\documentstyle[11pt]{article}
\begin{document}
\newcommand{\ECM}{\em Departament d'Estructura i Constituents de la
Mat\`eria
                  \\ Facultat de F\'\i sica, Universitat de Barcelona \\
                     Diagonal 647, E-08028 Barcelona}
\newcommand{\NPB}[3]{{\em Nucl.Phys.} {\bf B#1}(19{#2}){#3}}
\newcommand{\PRD}[3]{{\em Phys.Rev.} {\bf D#1}(19{#2}){#3}}
\newcommand{\PLB}[3]{{\em Phys.Lett.} {\bf B#1}(19{#2}){#3}}
\newcommand{\PRL}[3]{{\em Phys.Rev.Lett.} {\bf #1}(19{#2}){#3}}
\newcommand{\ZFP}[3]{{\em Zeit.Phys.} {\bf C#1}(19{#2}){#3}}
\newcommand{\PR}[3]{{\em Phys.Rev.} {\bf #1}(19{#2}){#3}}
\newcommand{\AP}[3]{{\em Ann. of Phys. (N.Y.)} {\bf #1}(19{#2}){#3}}
\newcommand{\IJA}[3]{{\em Int.J.Mod.Phys.} {\bf A#1}(19{#2}){#3}}
\newcommand{\CMP}[3]{{\em Commun.Math.Phys.} {\bf #1}(19{#2}){#3}}

\def\thefootnote{\fnsymbol{footnote}}
\pagestyle{empty}
{\hfill \parbox{6cm}{\begin{center} UB-ECM-PF 91/8\\
                                    October 1991
                     \end{center}}}
\vspace{1.5cm}

\begin{center}
\large{\bf DIFFERENTIAL RENORMALIZATION OF THE\\ WESS-ZUMINO MODEL}
\footnote{Work supported by Ministerio de Educaci\'on y Ciencia,
Madrid, Spain.} \end{center}
\vskip .6truein
\centerline {Peter E. Haagensen}

\vspace{.3cm}
\begin{center}
\ECM
\end{center}
\vspace{1.5cm}

\centerline{\bf Abstract}
\medskip

We apply the recently developed method of differential
renormalization to the Wess-Zumino model. From the explicit
calculation of a finite, renormalized effective action,
the $\beta$-function is
computed to three loops and is found to agree with previous existing
results. As a further, nontrivial check of the method, the
Callan-Symanzik equations are also verified to that loop order. Finally,
we argue that differential renormalization presents advantages
over other superspace renormalization methods, in that it avoids both
the ambiguities inherent to supersymmetric regularization by dimensional
reduction (SRDR), and the complications of virtually all other
supersymmetric regulators.
\newpage
\pagestyle{plain}
\def\QQa{\renewcommand{\baselinestretch}{1.1}\Huge\large\normalsize}
\def\a{\alpha}  \def\b{\beta} \def\g{\gamma} \def\G{\Gamma}
\def\d{\delta} \def\D{\Delta} \def\e{\epsilon} \def\ee{\varepsilon}
\def\z{\zeta} \def\th{\theta} \def\TH{\Theta} \def\tth{\vartheta}
\def\k{\kappa} \def\l{\lambda} \def\L{\Lambda} \def\m{\mu} \def\n{\nu}
\def\cs{\xi} \def\Cs{\Xi} \def\p{\pi} \def\P{\Pi} \def\r{\rho} \def\s{\sigma}
\def\S{\Sigma} \def\t{\tau} \def\y{\upsilon} \def\Y{\upsilon}
\def\f{\phi} \def\F{\Phi} \def\x{\chi} \def\ps{\psi} \def\Ps{\Psi}
\def\o{\omega} \def\O{\Omega} \def\vf{\varphi}
\def\pa{\partial} \def\da{\dagger} \def\dda{\ddagger}

\def\ph4{$\lambda\varphi^{4}$} \def\bfnc{$\beta$-function}
\def\dd{D_{\a}} \def\bdd{\bar{D}_{\dot{\a}}}
\def\dr{differential renormalization}
\def\vs{\vspace{.25in}}
\QQa

{\em 1. Introduction --} Since it has been developed, differential
renormalization has had some important successes. At considerable
calculational (and conceptual) simplification over other regularization
methods, it has been used to fully renormalize massless \ph4\ theory to
three loops[1], to calculate the triangle anomaly[1,2], and the 1-loop
\bfnc\
in Yang-Mills theory[1]. Further, mass effects in \ph4\ theory have also
been included straightforwardly[1,2], and the translation of the method
to a setting in which
counterterms appear explicitly has also been achieved in the course of
proving the unitarity of the method for massless \ph4\ theory[3]. Like
for the \ph4\ model, the perturbation theory of the Wess-Zumino model is
fairly simple and well-known, but at the same time sufficiently
nontrivial for it to represent a good testing ground for differential
renormalization, and we find that the method works as well in this case
as it does in \ph4\ theory.\\

The basic idea of the method is as follows: rather than regularizing by
altering the field content or the propagators of a theory (e.g.,
Pauli-Villars, higher derivatives, point splitting) or the space on
which it is defined (e.g., dimensional regularization), it regularizes
and renormalizes only the $x$-space amplitudes that are too divergent to
have a Fourier transform into momentum space. In practice, this is done
by writing amplitudes in $x$-space, expressing the divergent pieces as
derivatives of less singular terms, and then formally performing partial
integrations freely (which in fact will be precisely the implicit
regularization and renormalization that the method performs).
Counterterms do not appear at any point (although it is possible to keep
track of them explicitly[3]), and one goes directly from bare amplitudes
to renormalized ones, without a separate intermediate step of subtraction
in the regularized theory. The simplest example occurs at one loop both
for the 4-point function in \ph4\ theory and the 2-point function in the
Wess-Zumino model: these amplitudes will contain the term
$((x-x')^{2})^{-2}$, and its Fourier transform into momentum space (of
course, in the end one {\em must} go to momentum space to calculate
scattering amplitudes) is:\\

\[\int d^{4}\!x\, \frac{e^{ip\cdot x}}{x^{4}}=\frac{4\p^{2}}{p}
\int_{0}^{\infty}dr\, \frac{1}{r^{2}}J_{1}(pr)\, ,\]\\
where $r=(x^{2})^{\frac{1}{2}}$, and $J_{1}$ is the first-order Bessel
function. For $r\rightarrow\infty$ the integrand
converges fast enough, but for $r\rightarrow 0$, $J_{1}(pr)\sim
\frac{1}{2}pr$, and we have a logarithmic UV divergence. To regularize
it, differential renormalization then prescribes the use of the following
identity:

\begin{equation}
\frac{1}{x^{4}}=-\frac{1}{4}\Box\frac{\ln x^{2}M^{2}}{x^{2}}\, ,
\hspace{2cm}x\ne 0
\end{equation}\\

\noindent valid everywhere except at the origin. Now, the term on which
the D'Alembertian is acting is, as opposed to $\frac{1}{x^{4}}$, a
well-defined distribution with a proper Fourier transform. By formally
integrating by parts we are then able to define a regularized Fourier
transform for $\frac{1}{x^{4}}$:

\begin{eqnarray*}
\int d^{4}\!x\, e^{ip\cdot x}\frac{1}{x^{4}}&=&-\frac{1}{4}\int d^{4}\!x
\, e^{ip\cdot x}\Box\frac{\ln x^{2}M^{2}}{x^{2}}\equiv\frac{p^{2}}{4}
\int d^{4}\!x\, e^{ip\cdot x}\frac{\ln x^{2}M^{2}}{x^{2}}\\
&=&-\p^{2}\ln\frac{p^{2}}{\bar{M}^{2}}\, ,
\end{eqnarray*}

\noindent where $\bar{M}=\frac{2M}{\g_{E}}$, and $\g_{E}=1.78107...$ is
the Euler-Mascheroni constant. A general feature of differential
renormalization which can already be seen from the example above is that
in the process of expressing singular functions as derivatives of less
singular functions and then integrating by parts, thereby discarding
(infinite) surface terms, we are, on the one hand, removing the
divergence of an amplitude and, on the other, acquiring a new mass
parameter $M$ which appears in the guise of an integration constant.
Since the method does not keep track explicitly of the infinities being
subtracted, this mass parameter will be crucial for the study of the RG
invariance of renormalized amplitudes and the consistency of the method.\\

Because Eq.(1) is an identity except at the origin, bare and renormalized
amplitudes will also be identical except at isolated points. The
method does not, so to speak, uniformly and blindly regularize an entire
theory {\em ab initio} (like the regularization procedures mentioned
above), but rather removes divergences singly where they appear in
amplitudes. In this sense, differential renormalization is a ``minimal''
regularization: it alters the theory in the least amount possible, and in
particular it does not disturb its original symmetries.\\

To apply it to the Wess-Zumino model, we will use the manifestly
supersymmetric Feynman rules of Grisaru, Ro\v{c}ek, and Siegel[4,5]
(except for the fact that we do not go into momentum space), to
calculate loop corrections to the chiral propagator. For superspace
theories in general, using these Feynman rules, one ends up with the
following superspace effective action:

\begin{equation}
\G=\int d^{4}\!\th\sum_{N}\int d^{4}\! x_{1}...d^{4}\! x_{N}
F_{1}(x_{1},\th)
F_{2}(x_{2},\th)...F_{N}(x_{N},\th)G^{(N)}(x_{1},...,x_{N})\, ,
\end{equation}

\noindent where the $F_{i}(x_{i},\th)$ are functions of superfields and
their (supersymmetry) covariant derivatives, and $G^{(i)}$ are
translationally invariant functions of $i$ spacetime coordinates.
Because of the non-renormalization theorem[4,5], the Wess-Zumino model,
in particular, does not have a genuine coupling constant renormalization
(but only the one inherited from wavefunction renormalization)[6,7],
and is entirely renormalized by removing the divergences from
$G^{(2)}(x-x')$. That we do, up to three loops, with differential
renormalization; with more ease than previous methods, we find a
three-loop \bfnc\ which agrees with the existing results[7,8,9], and as
further consistency checks of the method, we verify the Callan-Symanzik
equation for $G^{(2)}_{ren}$, at each loop order, also to three loops.\vs

{\em 2. The Model and Loop Calculations -} The Wess-Zumino model is
described by the following superspace action:\\

\begin{equation}
S=\int d^{4}\!x \left(\int d^{2}\!\th d^{2}\!\bar{\th}\f\bar{\f}-
\frac{g}{3!}\left(\int d^{2}\!\th\f^{3}+\int d^{2}\!\bar{\th}\bar{\f}^{3}
\right)\right)\, ,
\end{equation}\\

\noindent where $(x^{a},\th^{\a},\bar{\th}^{\dot{\a}}),\, a=1,2,3,4,
\, \a =+,-,\, \dot{\a}=\dot{+},\dot{-}$ are coordinates of $d=4,\, N=1$
superspace, and $\f$ ($\bar{\f}$) is a chiral (antichiral) superfield:

\begin{eqnarray*}
\bdd\f&=&\left(\frac{\pa}{\pa\bar{\th}^{\dot{\a}}}+
i\th^{\b}\s^{a}_{\b\dot{\a}}\pa_{a}\right)\f(x,\th,\bar{\th})=0\\
\dd\bar{\f}&=&\left(\frac{\pa}{\pa\th^{\a}}+
i\s^{a}_{\a\dot{\b}}\bar{\th}^{\dot{\b}}\pa_{a}\right)\bar{\f}
(x,\th,\bar{\th})=0\, .
\end{eqnarray*}

\noindent In component form,

\[
\f (x,\th,\bar{\th})=\vf (x)+\th\ps (x)+\th\th
F(x)+i\th\s^{a}\bar{\th}\pa_{a}\vf (x)
-\frac{i}{2}\th\th\pa_{a}\ps (x)\s^{a}\bar{\th}+
\frac{1}{4}\th\th\bar{\th}\bar{\th}\Box\vf (x)
\]

\noindent and

\[
S=\int d^{4}\!x \left[-\pa_{a}\vf^{*}\pa^{a}\vf -\frac{i}{4}(
\ps\s^{a}\pa_{a}\bar{\ps} -\pa_{a}\ps\s^{a}\bar{\ps})+FF^{*}
-\frac{g}{3!}(\vf\vf F+\frac{1}{2}\ps\ps A+\, c.c.)\right]\, .
\]

For our calculations in superspace, we use the Feynman rules of Grisaru,
Ro\v{c}ek and Siegel[4,5]; in coordinate space, they are: $i$) the
propagator is

\begin{eqnarray*}
<\bar{\f}(x,\th)\f (x',\th ')>&=&\frac{1}{4\p^{2}(x-x')^{2}}
\d^{(4)}(\th -\th ')\\
&=&\frac{1}{4\p^{2}(x-x')^{2}}(\th -\th ')^{2}(\bar{\th} -\bar{\th}
')^{2}\, ;
\end{eqnarray*}

\noindent $ii$) for each vertex, include a factor of $\frac{g}{3!}$;
$iii$) for a chiral ($\f^{3}$) vertex, include two factors of
$-\frac{1}{4}\bar{D}^{2}$ acting on any two of the propagators for the
lines arriving at the vertex; for vertices containing an external line,
include only one $-\frac{1}{4}\bar{D}^{2}$ factor acting on any of the
two other (internal) lines; $iv$) same for antichiral vertices
($\bar{\f}^{3}$), with $-\frac{1}{4}D^{2}$ factors; $v$) integrate over
internal superspace points $(x_{int},\th_{int})$ and external $\th$'s,
$\th_{ext}$; $vi$) include appropriate symmetry factors for each
diagram. Finally, the $D$-algebra and transfer rules are:

\begin{eqnarray*}
D^{2}\bar{D}^{2}D^{2}=16\Box D^{2}\, &,&\,
\bar{D}^{2}D^{2}\bar{D}^{2}=16\Box\bar{D}^{2},\\
D^{3}=\bar{D}^{3}=0\, &,&\, [\dd ,\bar{D}^{2}]=4i\s^{a}_{\a\dot{\b}}
\pa_{a}\bar{D}^{\dot{\b}}\, ,\\
\dd (\th ',\pa_{x'})[\d^{(4)}(\th -\th ')f(x-x')]&=&-\dd (\th ,\pa_{x})
[\d^{(4)}(\th -\th ')f(x-x')]\, .
\end{eqnarray*}

With these rules for calculating Feynman diagrams, the effective action
takes the form (after all $\th$-integrations but the last have been
carried out):

\begin{eqnarray}
\G [\f,\bar{\f}]&=&\int d^{4}\!\th\left[\int d^{4}\! xd^{4}\! x'\f
(x,\th,\bar{\th})
\bar{\f}(x',\th ,\bar{\th})G^{(2)}(x-x')\right]\nonumber\\
&+&\left[\int d^{2}\!\th\int d^{4}\! x_{1} d^{4}\!
x_{2} d^{4}\! x_{3}\f(x_{1},\th ,\bar{\th})\f(x_{2},\th ,\bar{\th})
\f(x_{3},\th ,\bar{\th})G^{(3)}_{ch}(x_{1},x_{2},x_{3})+c.c.
\right]\nonumber\\
&+&\int d^{4}\!\th\,\left[\cdots \,\right]\, ,
\end{eqnarray}

where

\begin{eqnarray}
G^{(2)}(x-x')&=&\d^{(4)}(x-x')+\k_{1}(x-x')+\k_{2}(x-x')
+\k_{3}(x-x')+etc.\\
&&\nonumber\\
G^{(3)}_{ch}(x_{1},x_{2},x_{3})&=&-\frac{g}{3!^{2}}\left[\d^{(4)}
(x_{1}-x_{2})\d^{(4)}(x_{1}-x_{3})+\, 5\, perms.\right]
\end{eqnarray}

\noindent and the dots in Eq.(4) refer to higher-point functions which
are
either finite or are renormalized by the renormalization of $G^{(2)}$.
The diagrams contributing to $G^{(2)}$ are depicted in Figs.(1) and
(2). $\k_{1}$ is the one-loop contribution from Fig.(1.a), $\k_{2}$
the two-loop contribution, Fig.(1.b), and $\k_{3}$ contains
contributions from the four three-loop diagrams in Fig.(2).\\

The non-renormalization theorem determines that the renormalization of
$g$ comes solely from wavefunction renormalization, and dictates the
above form for $G^{(3)}_{ch}$, correct to all loops. By studying the
Callan-Symanzik equation for $G^{(3)}_{ch}$, we find a simple relaton
between the \bfnc\ and the anomalous dimension $\g (g)$ of $\f$:

\begin{eqnarray}
\left(M\frac{\pa}{\pa M}+\b (g)\frac{\pa}{\pa g}-3\g (g)\right)
G^{(3)}_{ch}(x_{1},x_{2},x_{3};g)&=&\nonumber\\
=\,\left(\b (g)\frac{\pa}{\pa g}-3\g (g)\right)
G^{(3)}_{ch}(x_{1},x_{2},x_{3};g)&=&0\nonumber
\end{eqnarray}
\begin{equation}
\Rightarrow \b (g)=3g\g (g)\, .
\end{equation}

The Callan-Symanzik equation for the renormalized 2-point function
then becomes:

\begin{eqnarray}
\left(M\frac{\pa}{\pa M}+\b (g)\frac{\pa}{\pa g}-2\g (g)\right)
G^{(2)}_{ren}(x-x';g,M)&=&\nonumber\\
\left(M\frac{\pa}{\pa M}+\b (g)\left(\frac{\pa}{\pa
g}-\frac{2}{3g}\right)\right) G^{(2)}_{ren}(x-x';g,M)&=&0\, .
\end{eqnarray}

In general, $M$ is a renormalization mass scale that governs the RG flow
of the theory; in  differential renormalization this scale appears
through the process described in the introduction.\\

We now proceed to renormalize the theory. For the diagrams of Figs.(1)
and (2) the
$D$-algebra and $\th$-integrations are straightforward, if tedious, and
we present only the final $x$-space expressions to be renormalized:

\begin{eqnarray}
\k_{1}(x-x')&=&\frac{1}{2}\left(\frac{g}{4\p^{2}}\right)^{2}
\frac{1}{(x-x')^{4}}\\
\k_{2}(x-x')&=&-\frac{1}{2}\left(\frac{g}{4\p^{2}}\right)^{4}
\frac{1}{(x-x')^{2}}\int d^{4}\!x_{1}\frac{1}{(x-x_{1})^{2}}
\frac{1}{(x_{1}-x')^{4}}\\
\k_{3a}(x-x')&=&\frac{1}{8}\left(\frac{g}{4\p^{2}}\right)^{6}
\left(\int d^{4}\!x_{1}\frac{1}{(x-x_{1})^{4}}\frac{1}{(x_{1}-x')^{2}}
\right)^{2}\\
\k_{3b}(x-x')&=&\frac{1}{2}\left(\frac{g}{4\p^{2}}\right)^{6}
\frac{1}{(x-x')^{2}}\times\nonumber\\
&&\int d^{4}\!x_{1} \int d^{4}\!x_{2}
\frac{1}{(x-x_{1})^{4}}\frac{1}{(x_{1}-x_{2})^{2}}
\frac{1}{(x-x_{2})^{2}}\frac{1}{(x_{2}-x')^{2}}\\
\k_{3c}(x-x')&=&\frac{1}{4}\left(\frac{g}{4\p^{2}}\right)^{6}
\frac{1}{(x-x')^{2}}\times\nonumber\\
&&\int d^{4}\!x_{1} \int d^{4}\!x_{2}
\frac{1}{(x-x_{1})^{2}}\frac{1}{(x_{1}-x_{2})^{4}}
\frac{1}{(x_{2}-x')^{4}}\\
\k_{3d}(x-x')&=&\frac{1}{2}\left(\frac{g}{4\p^{2}}\right)^{6}
\frac{1}{(x-x')^{2}}\times\nonumber\\
&&\int d^{4}\!x_{1} \int d^{4}\!x_{2}
\frac{1}{(x-x_{1})^{2}}\frac{1}{(x'-x_{2})^{2}}
\frac{1}{(x_{1}-x_{2})^{2}}\frac{1}{(x'-x_{1})^{2}}
\frac{1}{(x-x_{2})^{2}}\nonumber\\
 &=&\frac{1}{2}\left(\frac{g}{4\p^{2}}\right)^{6}
  6\p^{4}\z (3)\frac{1}{(x-x')^{4}}
\end{eqnarray}\\

This last (finite) integral was done with the aid of the Gegenbauer
polynomial technique [1,7]. Now, in the course of performing the
above integrals, we use the following differential renormalization
identities[1] whenever singular terms appear (i.e., terms like the
l.h.s. below):

\begin{eqnarray}
\frac{1}{x^{4}}&=&-\frac{1}{4}\Box\, \frac{\ln x^{2}M^{2}}{x^{2}}
\, , \hspace{.5cm}x\ne 0\\
\frac{\ln x^{2}M^{2}}{x^{4}}&=&-\frac{1}{8}\Box\,
 \frac{\ln^{2} x^{2}M^{2}+2\ln x^{2}M^{2}}{x^{2}}
\, ,\hspace{.5cm}x\ne 0\\
\frac{\ln^{2}x^{2}M^{2}}{x^{4}}&=&-\frac{1}{12}\Box\,
\frac{\ln^{3}x^{2}M^{2}+3\ln^{2}x^{2}M^{2}+6\ln x^{2}M^{2}}{x^{2}}
\, , \hspace{.5cm}x\ne 0\, .
\end{eqnarray}\\

\noindent With this, we are able to calculate the renormalized values of
$\k_{i}$:

\begin{eqnarray}
\k^{ren}_{1}(x)&=&-2g^{2}\left(\frac{1}{4\p}\right)^{4}
\Box\, \frac{\ln x^{2}M^{2}}{x^{2}}\\
\k^{ren}_{2}(x)&=&g^{4}\left(\frac{1}{4\p}\right)^{6}\Box\,
\frac{\ln^{2} x^{2}M^{2}+2\ln x^{2}M^{2}}{x^{2}}\\
\k^{ren}_{3a}(x)&=&-\frac{1}{6}g^{6}\left(\frac{1}{4\p}\right)^{8}\Box\,
\frac{\ln^{3}x^{2}M^{2}+3\ln^{2}x^{2}M^{2}+6\ln
x^{2}M^{2}}{x^{2}}\\
\k^{ren}_{3b}(x)&=&-g^{6}\left(\frac{1}{4\p}\right)^{8}\Box\,
\frac{\frac{1}{3}\ln^{3}x^{2}M^{2}+2\ln^{2}x^{2}M^{2}+4\ln
x^{2}M^{2}}{x^{2}}\\
\k^{ren}_{3c}(x)&=&-\frac{1}{3}g^{6}\left(\frac{1}{4\p}\right)^{8}\Box\,
\frac{\ln^{3}x^{2}M^{2}+3\ln^{2}x^{2}M^{2}+6\ln
x^{2}M^{2}}{x^{2}}\\
\k^{ren}_{3d}(x)&=&-12\z (3)g^{6}\left(\frac{1}{4\p}\right)^{8}
\Box\, \frac{\ln x^{2}M^{2}}{x^{2}}\, ,
\end{eqnarray}\\

\noindent where we have set $x'=0$ for simplicity. To calculate the
\bfnc\ and verify the consistency of the method we apply the
Callan-Symanzik equation, Eq.(4), to $G^{(2)}_{ren}=\d^{(4)}+
\k_{1}^{ren}+
\k_{2}^{ren}+\k_{3}^{ren}$. We write

\[\b (g)=\b_{1}g^{3}+\b_{2}g^{5}+\b_{3}g^{7}\]\\
and separate the pieces in the Callan-Symanzik equation
according to order in $g^{2}$ and the coefficients of the $\d$-function
and the different powers of log. All of these have to vanish
separately. At $O(g^{2})$, we get the one-loop \bfnc :

\begin{equation}
\b_{1}g^{3}=\frac{3}{2}\left(\frac{1}{4\p}\right)^{2}g^{3}\, .
\end{equation}

At $O(g^{4})$, the coefficient of $\d (x)$ vanishes for the following
value of the two-loop \bfnc :

\begin{equation}
\b_{2}g^{5}=-\frac{3}{2}\left(\frac{1}{4\p}\right)^{4}g^{5}\, ,
\end{equation}\\

\noindent and the coefficient of $\Box\frac{\ln x^{2}M^{2}}{x^{2}}$
vanishes for
the above value of $\b_{1}$; this is a consistency check of the method.
At  $O(g^{6})$, the vanishing of the coefficient of $\d (x)$ gives the
following three-loop \bfnc :

\begin{equation}
\b_{3}g^{7}=\left(\frac{21}{4}+9\z
(3)\right)\left(\frac{1}{4\p}\right)^{6} g^{7}\, .
\end{equation}\\

Finally, the coefficients of $\Box\frac{\ln x^{2}M^{2}}{x^{2}}$ and
$\Box\frac{\ln^{2}x^{2}M^{2}}{x^{2}}$ will vanish for the above values
of $\b_{1}$ and $\b_{2}$, and these are then two further consistency
checks. Our final result for the \bfnc\ is then:

\begin{equation}
\b (g)=g\left[\frac{3}{2}\left(\frac{g}{4\p}\right)^{2}-
  \frac{3}{2}\left(\frac{g}{4\p}\right)^{4}+
  (\frac{21}{4}+9\z (3))\left(\frac{g}{4\p}\right)^{6}\right]\, .
\end{equation}\\

We note that although the three-loop part is scheme-dependent, as
opposed to the one- and two-loop results, the coefficient of $\z (3)$
should be universal (barring, of course, a coupling constant
redefinition involving $\z (3)$) because the only diagram at three
loops leading to the transcendental function $\z (3)$ is primitively
divergent. While our coefficient of $\z (3)$ coincides with that
obtained in [9], it does not agree with [7]. The correct result is
that of [9], however[10]; in fact, because that is a four
loop calculation, it is possible to verify the three-loop result by RG
pole equations (which the authors of [9] in fact do).\vs

{\em Conclusions -} We have seen that the method of \dr\ is readily
applicable to the superspace Wess-Zumino model and, in particular, the
\bfnc\ calculations were considerably simpler than in SRDR [7,9].
Firstly, the integrations performed here were either trivial
(integrations with $\d$-functions) or very simple convolutions;
secondly, we did not have to keep track of subtraction of
subdivergences: at each loop order the renormalization is done in a
single step, by the direct use of ``\dr\ identities'' like
Eqs.(12)-(14)
whenever singular expressions are encountered (in SRDR, on the other
hand (cf. Eq.(2.11) of [7]), a careful account needs to be kept of
different renormalization constants corresponding to contributions of
subdivergences at higher loops).\\

We have attempted to describe the differential renormalization of the
Wess-Zumino model in such a way that the extension to other superspace
theories becomes obvious:
one should again
start with the general form for the effective action, Eq.(2), with the
$G^{(N)}$ written in terms of $x$-space integrals, and apply
differential renormalization identities to these integrals, thus
regularizing and renormalizing the theory.
Differential renormalization will take translation-invariant quantities
$G^{(N)}$
into translation-invariant quantities $G^{(N)}_{ren}$, and
will manifestly maintain supersymmetry.
Naturally, we expect
this procedure to be far simpler than several other existing
regularization methods (e.g., the supersymmetric versions of
Pauli-Villars, higher derivative and point-splitting methods).
Furthermore, if the method proves to be as simple to implement in
general as SRDR,
it would have the advantage that the ambiguities associated with SRDR
would be avoided: like for dimensional regularization, SRDR presents
ambiguities due to the presence of intrinsically four-dimensional
quantities, like $\ee^{\m\n\r\s}$, in a dimensionally continued setting
[5,11]. Differential renormalization, of course, never leaves four
dimensions and would thus avoid these problems altogether. One initial
indication of the further applicability of the method is the fact that
gauge (vector) superfield propagators are identical to the massless
chiral propagators we have considered here. Thus, calculations with
vector superfields should not in principle present any new difficulty;
for instance, the one-loop correction to the vector
propagator in supersymmetric Yang-Mills theory coupled to a massless
chiral superfield
is again as simple to compute as in SRDR.
Further work along these lines is in progress.\vs

{\em Acknowledgments -} It is a pleasure to thank J.I. Latorre for
always illuminating discussions. Comments on the manuscript from D.Z.
Freedman and J.I. Latorre are also very appreciated.\vs

{\em References}\\

\noindent
{\bf [1]} Freedman, D.Z., K. Johnson and J.I. Latorre, to be published
in {\em Nucl.Phys.} {\bf B}.\\
{\bf [2]} Freedman, D.Z., Proceedings of the Stony Brook Conference on
Strings and\\ \mbox{}\phantom{[1] }Symmetries, Spring 1991.\\
{\bf [3]} Freedman, D.Z., R. Mu\~noz-Tapia, X.
Vilas{\'{\i}}s-Cardona, manuscript in preparation.\\
{\bf [4]} Grisaru, M.T., M. Ro\v{c}ek and W. Siegel,
\NPB{159}{79}{429}.\\
{\bf [5]} Gates, J., M.T. Grisaru, M. Ro\v{c}ek and W. Siegel, {\em
Superspace}, Benjamin-Cummings,\\ \mbox{}\phantom{[1] } 1983.\\
{\bf [6]} Iliopoulos, J. and B. Zumino, \NPB{76}{74}{310}.\\
\mbox{}\phantom{[1] }Ferrara, S., J. Iliopoulos and B. Zumino,
\NPB{77}{74}{413}.\\
{\bf [7]} Abbott, L.F. and M.T. Grisaru, \NPB{169}{80}{415}.\\
{\bf [8]} Townsend, P.K. and P. van Nieuwenhuizen, \PRD{20}{79}{1832}.\\
{\bf [9]} Sen, A. and M.K. Sundaresan, \PLB{101}{81}{61}.\\
{\bf [10]} M.T. Grisaru, private communication.\\
{\bf [11]} Siegel, W., \PLB{94}{80}{37}.\vs

{\em Figures}\vspace{8cm}

\centerline{Fig.(1): One- and two-loop contributions to
$G^{(2)}(x-x')$.}
\vspace{8cm}

\centerline{Fig.(2): Three-loop contributions to
$G^{(2)}(x-x')$.}
\end{document}